# From diluted magnetic semiconductors to self-organized nanocolumns of GeMn in Germanium

S. Tardif [a], Ing-Song Yu [b], T. Devillers [b], M. Jamet [b], S. Cherifi [a],
J. Cibert [*a], A. Barski [b], P. Bayle-Guillemaud [b], E. Bellet-Amalric [b],

[a] Institut Néel, CNRS-UJF, BP166, 38042 Grenoble, France
[b] INAC, CEA-Grenoble, 38054 Grenoble, France

## ABSTRACT

While achieving high Curie temperatures (above room temperature) in diluted magnetic semiconductors remains a challenge in the case of well controlled homogeneous alloys, several systems characterized by a strongly inhomogeneous incorporation of the magnetic component appear as promising. Incorporation of manganese into germanium drastically alters the growth conditions, and in certain conditions of low temperature Molecular Beam Epitaxy it leads to the formation of well organized nanocolumns of a Mn-rich material, with a crystalline structure in epitaxial relationship with the Mn-poor germanium matrix. A strong interaction between the Mn atoms in these nanocolums is demonstrated by x-ray absorption spectroscopy, giving rise to a ferromagnetic character as observed through magnetometry and x-ray magnetic circular dichroism. Most interesting, intense magneto-transport features are observed on the whole structure, which strongly depend on the magnetic configuration of the nanocolumns.

## 1. INTRODUCTION

Making ferromagnetic a semiconductor[1] compatible with mainstream silicon technology, like germanium, is highly desirable. Following early reports[2] on $Ge_{1-x}Mn_x$ as a homogeneous diluted magnetic semiconductor, recent reports have focused on a strongly inhomogeneous incorporation of Mn in germanium[3]. A dramatic example is that of Mn-rich self-organized nanocolumns running throughout the germanium layer[4]. The nanocolumns have the same structure as the matrix. They exhibit a large ferromagnetic interaction, and strongly influence the magneto-transport properties of the whole layer, with e.g. a strong anomalous Hall effect (Hall angle as high as 0.6 rad at low temperature) persisting up to above room temperature.

Such inhomogeneous systems[5] appear as quite promising as they may open a way towards an all-semiconductor spintronics operating at room-temperature. They thus offer an alternative to homogeneous diluted magnetic semiconductors such as (Ga,Mn)As - for which rising the critical temperature above the presently achieved value (now over 180 K) is still a challenge for epitaxy - or wide bandgap semiconductors which are now recognized as being governed by more complex mechanisms[6] than the RKKY-like ferromagnetism induced by free carriers.

Here we will first summarize recent experimental results obtained on homogeneous diluted magnetic semiconductors, and particularly those[7] based on wide bandgap semiconductors such as GaN. Then we will turn to GeMn and discuss the influence of the growth conditions (growth temperature, Mn concentration, substrate morphology …) on the nature of the Mn-rich inclusions and on the morphology of the Mn-rich nanocolumns. We will discuss the electronic state of the Mn atoms in the nanocolumns, at the light of x-ray absorption spectroscopy (XAS) and x-ray magnetic circular dichroism (XMCD) obtained[8] with and without a capping layer. Finally, we will show how the structural properties of our GeMn layers and the morphology of the nanocolumns strongly influence[9] their electronic and magnetic properties.

---

[*] *joel.cibert@grenoble.cnrs.fr; http://neel.cnrs.fr

## 2. FERROMAGNETISM IN HOMOGENEOUS DILUTED MAGNETIC SEMICONDUCTORS AT A GLANCE

Most of the interest in semiconductor physics is related first to our ability to dope the materials with electrically active impurities, but also and perhaps even more, to the development of epitaxial growth, which allows us to combine different semiconductors into heterostructures and nanostructures. In these, interfaces confine the electrons and holes, which allow us to manipulate these charge carriers using electric fields or light pulses.

In such semiconductors, introducing magnetic impurities in substitutional sites creates a diluted magnetic semiconductor (DMS), and such a doping with magnetically active impurities brings yet another degree of freedom, essentially due to the strong coupling between the localized spins of the impurities and the carriers in the band of the semiconductor (the so-called giant Zeeman effect). In the presence of a sufficiently high density of holes, several DMSs have been shown to become ferromagnetic. Such materials are fascinating from the viewpoint of basic research, but applications require that the critical temperature be above room temperature - while keeping all specific features of the host semiconductor.

In spite of huge efforts to increase the critical temperature of homogeneous DMSs at, or preferably well above, room temperature, this remains a challenge. If we consider only "classical" semiconductors (by this we mean semiconductors of the column IV, or III-Vs or II-VIs, with the diamond, zinc blende or wurzite structure), the best documented ferromagnetic DMS is $Ga_{1-x}Mn_xAs$. For a long time, the critical temperature was at most 110 K[1]. Then the presence of interstitial Mn was identified as the origin of this limitation[10,11]. Now, in samples where interstitial Mn have been eliminated as much as possible by proper thermal annealing[12], the critical temperature increases quite linearly with the density of substitutional Mn up to $T_c \approx 188$ K. Even if the linear increase is encouraging, there is still a long way towards carrier induced ferromagnetism at room temperature.

Another model system is based on tellurides: in the 3D form in ZnMnTe[13], or in the 2D form in CdMnTe[14]. In both cases extra doping is needed to bring the holes, since Mn is an isoelectronic center in II-VIs. The critical temperature remains much lower in tellurides than in $Ga_{1-x}Mn_xAs$. This was understood[13,15] by taking into account properly the structure of the valence band and the variation of the spin-hole exchange energy with lattice parameter. To be short, a stronger spin-orbit coupling in the tellurides, implies that the spin of the hole is aligned on the **k**-vector, so that a large part of the carriers do not take part in the spin-spin coupling. In addition, a decrease of the spin-hole exchange energy when the lattice parameter increases was evidenced by two independent methods: a direct measurement using the giant Zeeman splitting of the exciton, on one hand, and a calculation using the band parameters determined by x-ray spectroscopy, on the other hand[15]. Such considerations lead to the idea that[16] wide bandgap semiconductors, with a small spin-orbit coupling and a small lattice parameter, could present ferromagnetic properties above room temperature, provided they contain a large density of Mn impurities with the $d^5$ configuration together with a large density of holes.

Many studies - theoretical and experimental - were conducted on wide bandgap semiconductors with magnetic impurities. The picture that emerged progressively[17] is that two assumptions made in the above prediction are not verified: (1) contrary to what happens in GaAs, where it behaves as an acceptor and assumes the $d^5$ configuration, the Mn impurity in GaN is not an acceptor and the d-shell is less than half-filled[18]; (2) in the presence of a very strong coupling with the localized spins[19], such as realized in wide bandgap semiconductors like GaN and ZnO, the holes states split into bound states (which are localized on the magnetic impurity) and extended states spin-hole exchange energy (as those observed in magneto-optical spectroscopy, which results in a small giant Zeeman effect[20]).

Actually, recent evidences of high critical temperatures in semiconductors containing magnetic impurities have been ascribed to inhomogeneous materials. A good example is (Zn,Cr)Te, which exhibits high values of $T_c$ and at the same time strong magneto-optical effects[21,22]. In this material, Transmission Electron Microscopy evidences the presence of Cr-rich nanocrystals, and co-doping with electrically active impurities dramatically changes the magnetic properties through a change in the morphology of these nanocrystals[23].

Thus we have - at least - three ways to explore:

- DMS with a moderate bandgap, such as GaAs or closely related materials, for which there is no experimental evidence of a saturation of $T_c$ if one correctly controls the introduction of an increasing quantity of Mn impurities[12];

- wide bandgap DMS, which exhibit a localization of the strongly coupled holes - a mechanism which could be significantly screened if a high density of carriers could be introduced, but this is a formidable challenge for epitaxy[19];

- inhomogeneously doped semiconductors[24,25], with at the same time a strong interaction between the magnetic species in the region where their density is high, and a strong interaction with the carriers in the rest of the crystal, giving rise to interesting magneto-optical and magneto-transport properties.

We describe now the results obtained on germanium epilayers with manganese. This system belongs to the third category: indeed GeMn films contain nanocolumns of a Mn-rich GeMn compound, and exhibit a very strong anomalous Hall effect.

## 3. GERMANIUM IN MANGANESE

The rest of this paper aims at giving an overview - albeit oversimplified - of the main characteristics of (Ge,Mn) layers grown by molecular beam epitaxy and featuring Mn-rich nanocolumns. Well formed nanocolumns are obtained in samples grown in a rather narrow temperature range around 130°C. They run over the entire layer, exhibit a ferromagnetic character up to above room temperature which interests the whole nanocolumn, and they deeply affect the magneto-transport properties of the layer. At a lower growth temperature, the columns are smaller in diameter and part of them break into several segments; as a result, a superparamagnetic behaviour is observed. Samples grown at higher temperature show $Ge_3Mn_5$ clusters with a superparamagnetic behavior and a weak influence on the carriers in the layer.

More details are given in Ref. [4, 8, 9, 26].

### 3.1 Growth and morphology

Samples were grown by molecular beam epitaxy using two Knudsen effusion cells containing Ge and Mn. The deposition rate was kept low, about 0.2 Å s$^{-1}$. We used mostly epiready Ge (001) substrates, with a 40 nm-thick Ge buffer layer grown at 250 °C after thermal desorption of the surface oxide. The 80-nm-thick $Ge_{1-x}Mn_x$ layers were grown at low substrate temperature (80 to 200 °C): Mn has a remarkable surfactant effect. The average Mn content was adjusted between 1 and 11%, as determined by x-ray fluorescence performed on thick samples and by Rutherford backscattering. We focus here on samples with about 6% Mn, more information on the dependence on the Mn content can be found in Ref. [26]. A few samples have been grown on (001) GaAs substrates, either bare epiready substrates, or GaAs substrates with a GaAs buffer layer capped with amorphous Arsenic. More details are given in Ref. [4, 9, 26].

The most dramatic feature is the presence, in samples grown around 130°C, of nanocolumns running over the whole layer, as shown in Fig. 1a by transmission electron microscopy (TEM). On high-resolution plane views, Fig. 1b, these nanocolumns display the same crystalline structure as the Ge matrix, with often two dislocations in opposition on one diameter (Fig. 1c and d). The presence of these dislocations reveals that the lattice parameter of the material which constitutes the nanocolumns is larger than that of Ge.

Electron Energy Loss Spectroscopy (see Ref. 4) shows that these nanocolumns are Mn-rich. A more complete analysis of the Electron Energy Loss Spectroscopy images involving the average diameter of these columns (typically 3 nm), their areal density, and the Mn density averaged over the whole layer, suggests that their composition is close to $Ge_2Mn$ while the Mn content in the surrounding Ge matrix is very low, 1% at most.

In samples grown at lower temperature (below 120°C), the nanocolumns are smaller in diameter, and fully strained to the Ge lattice, see Ref. 4. There is some evidence on the electron microscopy cross-section images that they tend to break into pieces, a trend which will be confirmed by the study of magnetization, see below. Elongated, Mn-rich nano-crystals, which do not cross the entire layer, have been observed by other groups [27, 28].

In samples grown at higher temperature (typ. 150°C), TEM shows a strong disorder in the core of these nanocolumns: it is however difficult to decide whether an amorphous core was present in the as-grown layer, or has been caused by the thinning process. When increasing the growth temperature, $Ge_3Mn_5$ nanoclusters progressively appear. These clusters have a different crystalline structure. In samples grown at 180°C, where they are the dominant feature, their presence is easily detected from the RHEED pattern. They also characterize samples which have been annealed (we observe the $Ge_3Mn_5$ phase in samples annealed at 650°C). Such $Ge_3Mn_5$ clusters have been observed by other groups in as-grown samples[29,30] and their magnetic properties are well documented.

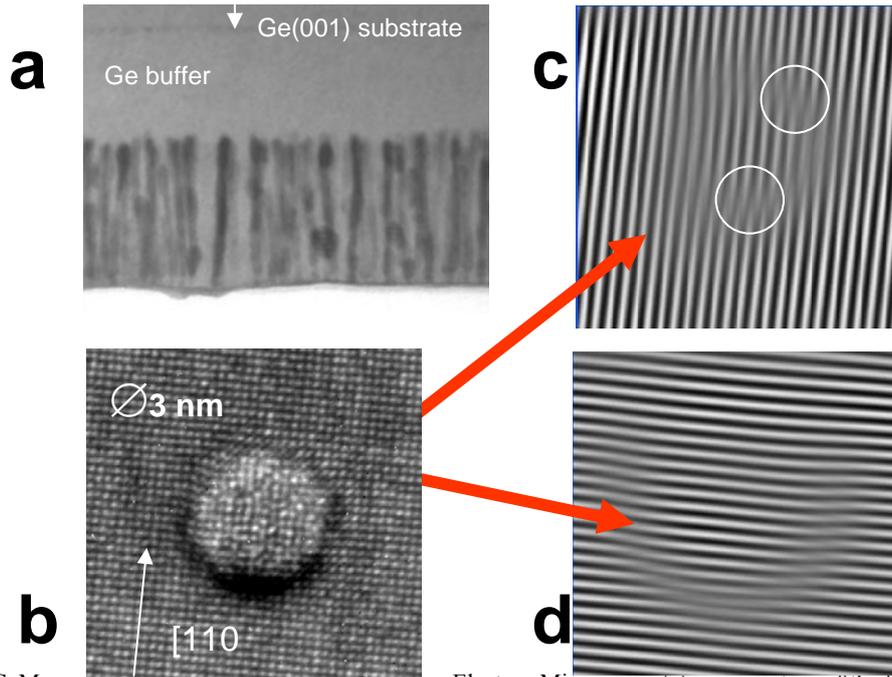

Figure 1: GeMn nanocolumns observed by Transmission Electron Microscopy (a) cross section of the 80 nm thick (Ge,Mn) layer, with the nanocolumns appearing as dark lines; (b) high resolution plane view; (c) Bragg filtering of the (2-20) spot evidencing a pair of dislocations; (d) Bragg filtering of the (220) spot with no dislocation (adapted from Ref. 4).

### 3.2 X-ray absorption spectroscopy and dichroism

X-ray absorption spectroscopy (XAS) and x-ray magnetic circular dichroism (XMCD) have been measured[8] at the Mn $L_{2,3}$-edges, at the synchrotron radiation facility BESSY. The samples were cooled down to 5 K and the x-ray absorption and dichroism have been measured in total electron yield mode with a magnetic field up to 6 T applied in the propagation direction of the photon beam. The typical absorption spectra are given in Fig. 2, for two samples with the same characteristics of the nanocolumns (layer thickness 60 nm, column diameter 3 nm and column density $3\times10^4$ µm$^{-2}$): The first sample is uncapped and the second one is capped with a germanium-silicon amorphous layer.

Fig. 2(a) shows XAS and XMCD spectra recorded from the uncapped sample. They exhibit sharp multiplet structures that are very similar to spectra recorded from ionic systems such as manganese oxide. However, the presence of the corresponding features in the XMCD spectra measured over at the same energy range indicate that the system is not (or at least not entirely) antiferromagnetic, as would be MnO at this temperature [8]. Note also that spectra exhibiting similar multiplet structures, albeit weaker, have been attributed to isolated Mn atoms at the surface of GaAs or Ge[31] or diluted in a DMS such as $Ga_{1-x}Mn_xAs$ [32].

The spectra recorded from a sample efficiently capped against oxidation (Fig. 2(b)) are strikingly different. Such spectra show a single-peak structure that is typical for metals, with broadened d-states. Note also that spectra from a (Ge,Mn) layer also exhibit such broadened spectra[35]. In our capped samples, the intensity of the XMCD $L_3$ peak closely follows[8] the magnetization measured by SQUID. We could attribute this magnetic signal without ambiguity to the $Ge_2Mn$ nanocolumns and rule out any contribution from parasitic $Ge_3Mn_5$ clusters.

The typical absorption spectra of metallic $Ge_3Mn_5$ clusters are quite similarly broadened[30]. Nevertheless, the application of the usual theoretical sum rules to the XMCD data concludes to a total magnetic moment of 2.34 µ$_B$ in $Ge_3Mn_5$ while a two times smaller value is measured in our capped samples. In both cases a very small orbital contribution to the magnetic moment is observed – typical of a half-filled d shell ($d^5$ configuration).

In summary, our study reveals that: (1) the sharp multiplet structures shown in Fig. 2(a) are due to oxidized GeMn, while the broad spectra in the same figure characterizes the $Ge_2Mn$ material which forms the nanocolumns, and (2) a strong broadening of the d-levels takes place in this material.

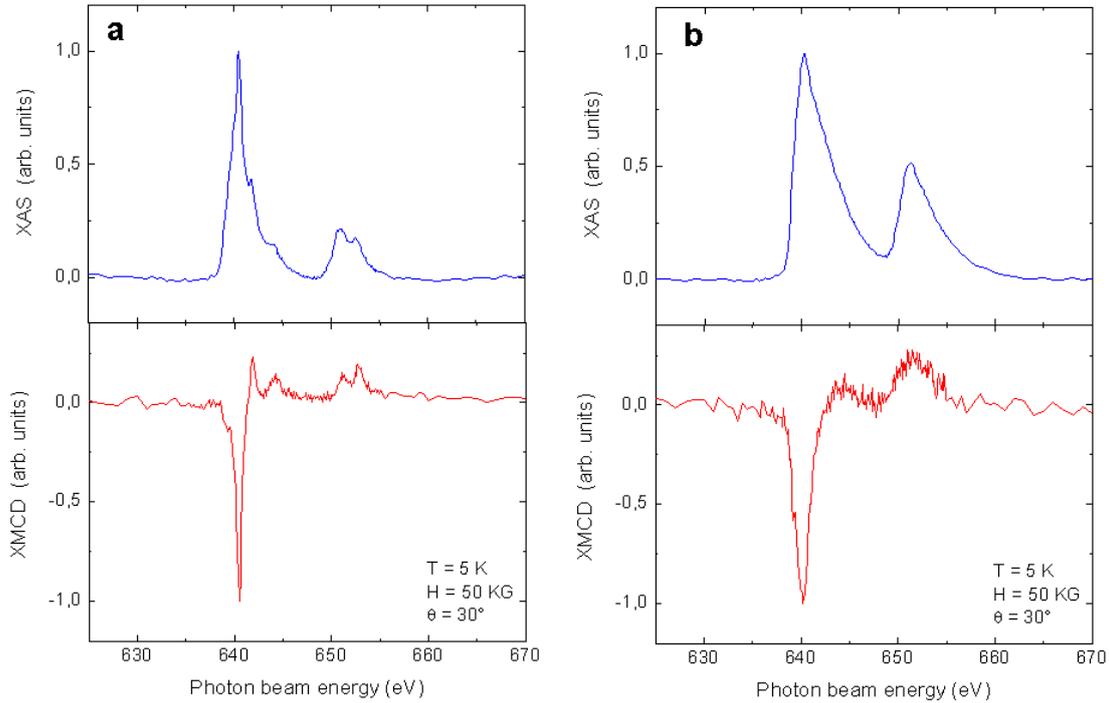

Figure 2: X-ray Absorption and dichroism spectra at the Mn $L_{2,3}$-edges, observed on two GeMn samples which differ only by the presence (b) or not (a) of a germanium-silicon cap layer. [adapted from Ref. 8].

### 3.3 Magnetic properties

As can be expected, the magnetic properties strongly depend on the nature and the morphology of the nanostructures embedded in the (Ge,Mn) layer. To summarize, we can observe four typical contributions: Mn atoms diluted in the Ge matrix, low-$T_C$ nanocolumns, high-$T_C$ nanocolumns, and $Ge_3Mn_5$ clusters. The relative weight of each phase principally depends on the growth temperature, and to a lesser extent on Mn concentration. We found no evidence of $Ge_8Mn_{11}$ which gives[33,34] a ferromagnetic signal between 150 and 300 K.

Fig. 3 gives, as typical examples, magnetization data for three samples grown at three different temperatures: in Fig. 3a, a sample grown at low temperature, in Fig.3b-c, a sample grown at 130°C, and in Fig. 3d, a sample grown at high temperature. More data and a complete analysis are found in Ref. 26. The inset of Fig. 3c shows the magnetization of an annealed sample where only $Ge_3Mn_5$ clusters subsist.

Let us start with the well-known example of $Ge_3Mn_5$ clusters as they are observed in the sample annealed at 650°C (inset of Fig. 3c). This figure shows the so-called FC-ZFC curves. In the ZFC measurement, the magnetization is recorded by cooling down the sample without applying the field, then applying the field at the lowest temperature, and warming up. In the FC measurement, the magnetization is recorded by applying the field at room temperature and cooling down the sample. The so-called blocking temperature sets the limit between a high temperature range (where the magnetization of a cluster is free to align along the applied field: the system is superparamagnetic and the FC and ZFC curves coincide) and a low-temperature range (where the clusters are magnetically frozen or ferromagnetic, so that the ZFC curve stays below the FC one). In the present case, both curves vanish around 300 K, which coincides with the value, 296 K, quoted[36] for the Curie temperature of bulk $Ge_3Mn_5$.

The magnetic signature of well formed nanocolumns is completely different. The magnetization persists up to above 400 K, with identical FC and ZFC curves (Fig. 3c). Thus the nanocolumns are ferromagnetic, with a critical temperature above 400 K. In addition to this ferromagnetic behavior, an increase of the magnetization below 50 K (which leads to a difference between the magnetization loops measured at different temperatures, Fig. 3b) is attributed to Mn diluted in the Ge matrix.

The magnetization of the sample grown at low temperature, Fig. 3a, exhibits FC-ZFC curves pointing to superparamagnetic objects with a low blocking temperature (about 15 K in this sample), and a Curie temperature below 170 K. In addition, the fit of the magnetization loops (not shown) suggests that the superparamagnetic moment is about 4 times smaller than the sum of the magnetic moments of the Mn atoms present in a columns, i.e., from a magnetic point-of-view, the nanocolumns break into ≈ 4 independent pieces.

Finally, the magnetization of the sample grown at higher temperature, Fig. 3d, can be understood as a superposition of the magnetization due to $Ge_3Mn_5$ clusters, with a blocking temperature close or equal to the Curie temperature ≈ 300 K (i.e., clusters are big enough to be ferromagnetic), and a contribution from nanocolumns (identified by the arrow in Fig. 3d).

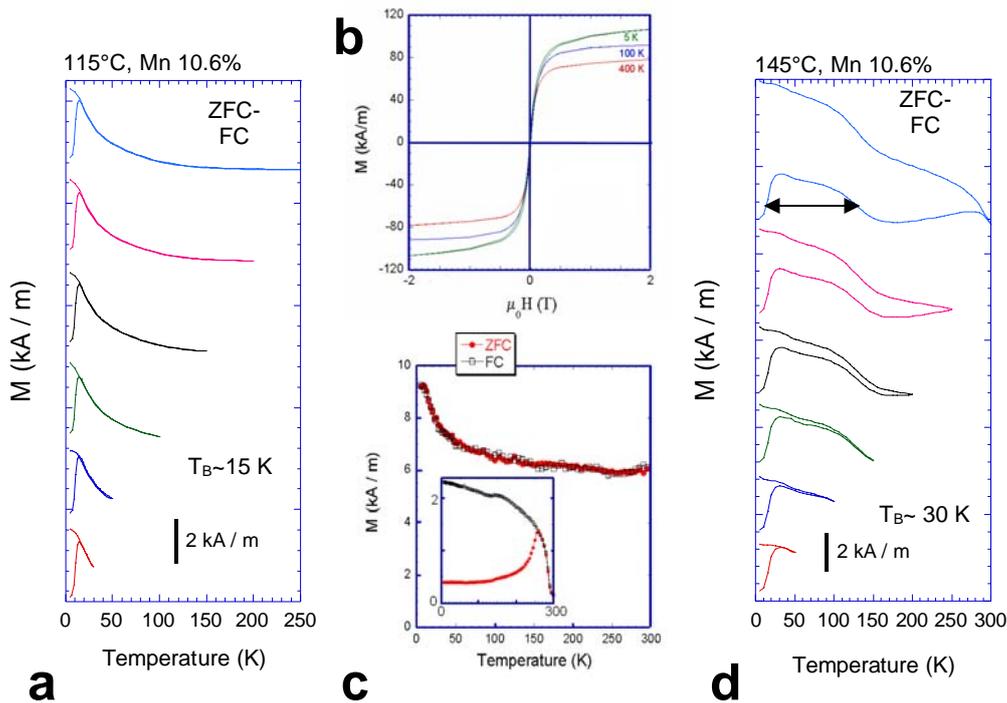

Figure 3: Magnetization curves. (a) FC-ZFC curves measured with a field of 15 mT applied in-plane, for a sample with 10.6% Mn grown at low temperature (115°C). The pairs of curves with different maximum temperatures are shifted for clarity. (b) Magnetization loops measured at 5 K, 100 K, 400 K, with the field applied in-plane, for a sample with 6% Mn grown at 130°C. (c) FC-ZFC curves measured with a field of 10 mT applied in-plane, for the same sample; inset shows the FC-ZFC curves measured after annealing at 650°C. (d) FC-ZFC curves measured with a field of 15 mT applied in-plane, for a sample with 10.6% Mn grown at high temperature (145°C). (adapted from Ref. 26)

To summarize, there is a competition for the incorporation of Mn under three forms:

- dilution of Mn atoms in the germanium matrix,

- formation of nanocolumns,

- formation of $Ge_3Mn_5$ clusters.

Increasing the growth temperature improves the continuity of the nanocolumns, and increases their diameter, so that their magnetization changes, from a superparamagnetic behavior of the nanocolumns or parts of them, to a ferromagnetic character of the whole nano-column. However a further increase of growth temperature also favors the formation of the $Ge_3Mn_5$ clusters, which finally take over at the expense of the nanocolumns.

### 3.4 Magneto-transport

The interest in DMSs is directly related to the magneto-transport or magneto-optical properties they may present, and to the possibility that these properties could be used as a basis for a semiconductor spintronics.

As far as magneto-transport is considered, the well-formed $Ge_2Mn$ nanocolumns are particularly attractive, even as an ensemble of nanocolumns embedded in a Mn-poor germanium layer. In particular, a very strong anomalous Hall effect is observed, as shown below.

The conductance of samples featuring well-formed nanocolumns was found to be *p*-type. This agrees with early observations[2] and could be caused by substitutional Mn atoms in Ge, which is expected to be a double acceptor (and actually a signature of diluted Mn atoms is seen on the magnetization). When the temperature is decreased from room temperature, the resistivity first decreases due to the suppression of phonon scattering, then a freezing takes place below 30 K, see Ref. 4 and 9.

Fig. 4a displays the Hall angle measured for different temperatures: a huge angle is measured, up to 0.6 at low temperature, and it persists up to room temperature. The value measured at low temperature is more than one order of magnitude larger than in conventional DMSs as $Ga_{1-x}Mn_xAs$ [Ref. 1] or $Zn_{1-x}Mn_xTe$ [Ref. 13]. Note that Fig. 4a shows quite a constant magnitude of the Hall angle for different temperatures below 30 K, while the resistivity increases by more than one order of magnitude in the same temperature range. In a homogeneous DMS, this would suggest[37] that the main mechanism could be skew scattering, and not the side-jump (or intrinsic mechanism) as in $Ga_{1-x}Mn_xAs$ [Ref. 38] or $Zn_{1-x}Mn_xTe$ [Ref. 13].

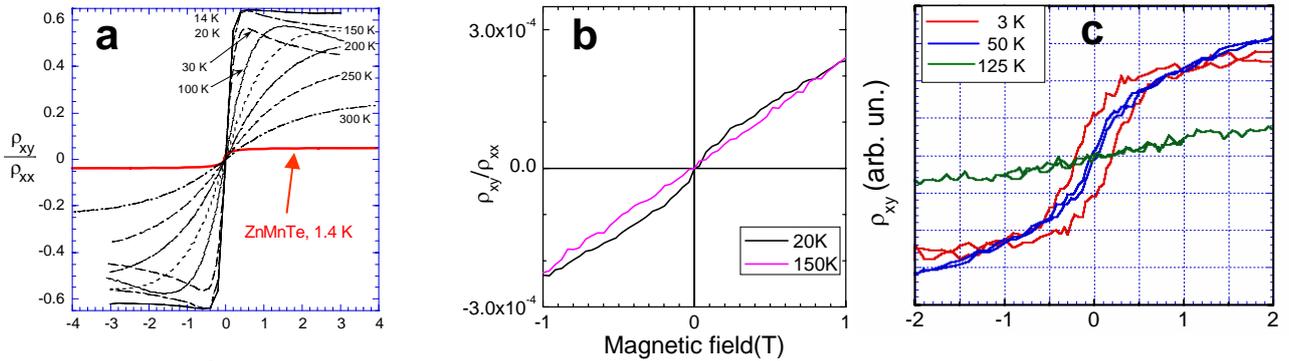

Figure 4 : Anomalous Hall Effect at different temperatures as indicated, as a function of the applied field, on two GeMn samples. (a) shows the data of a sample with well organized nanocolumns (see Fig. 1 and 3b). (b,c) shows the data for a sample with non parallel nanocolumns and a critical temperature about 110 K. In (a) and (b), the Hall angle (ratio of the transverse and longitudinal resistances) is shown. (c) shows the transverse resistance after subtraction of the linear contribution from ordinary Hall effect. The Hall angle measured on a homogeneous DMS (ZnMnTe) at low temperature is shown in (a) for comparison. [Adapted from Ref. 4, 9, 13].

The mechanism of this anomalous Hall effect however is still unknown - besides the fact that it demonstrates a strong interaction between the holes in the valence band of the Mn-poor germanium matrix, and the magnetic configuration of the Mn-rich nanocolumns. It is extremely sensitive to the microscopic details in the layer: for instance, we measured a very small Hall angle in a sample where the nanocolumns are not parallel (due to the rough surface of the GaAs substrate after thermal removal of the native oxide layer) and the carrier density is much smaller, see Fig. 4b. An optimization of the interaction is clearly needed.

## 3. CONCLUSIONS

The diagram describing the incorporation of manganese into a germanium layer grown by molecular beam epitaxy is quite rich, and the literature describes many different features observed at the nanometer scale. Our observations by RHEED, x-ray diffraction, electron microscopy, x-ray absorption and magnetic circular dichroism, magnetometry, and magneto-transport, require essentially three different objects to be understood:

- the incorporation of Mn-atoms, diluted into the Ge matrix, behaving as an acceptor (p-type conductivity) and giving rise to a paramagnetic component to the magnetization; this does not rule out a non-random distribution of these atoms.

-the formation of Mn-rich ($Ge_2Mn$) nanocolumns; in a narrow temperature range around 130°C, these nanocolumns run over the entire layer, the Mn atoms strongly interact within a column, and the columns are ferromagnetic with a high critical temperature; at lower growth temperature (presumably due to a small mobility) and at higher temperature (due to competition with the formation of $Ge_3Mn_5$ clusters), the nanocolumns break into short segments and they show superparamagnetism; when grown on rough substrates, the columns are not anymore parallel to each other. The nanocolumns strongly influence the magneto-transport properties of the whole layer.

- when the growth temperature is increased, $Ge_3Mn_5$ clusters progressively appear and take over the previous forms. They are also the dominant form in annealed samples. We did not measure any clear influence of the $Ge_3Mn_5$ clusters on the magneto-transport.

The main challenge is now to obtain a better control of the formation of the nanocolumns, including a further control of the different forms by codoping with electrically active impurities[24] and a control of their position, e.g., by patterning the substrate. The present observations suggest fascinating possibilities of spin manipulation in a single nanocolumn, or in well defined segments which have been fabricated by modulating the Mn flux during growth[26].

Nevertheless, the presence of an anomalous Hall effect which can be exceptionally large at low temperature and persists up to room temperature, already suggests many opportunities of using the magneto-transport properties of the whole layer, with the current propagating in the plane.

## ACKNOWLEDGEMENTS

We thank Nora Darowski and Detlef Schmitz for the support during the XAS and XMCD measurements at BESSY.